\documentclass[prb,aps]{revtex4}

\usepackage{graphicx}
\usepackage{epsfig}
\usepackage{dcolumn}
\usepackage{bm}

\begin{document}

\title{Current noise of a single-electron transistor coupled to a
nanomechanical resonator}

\author{A. D. Armour}
\affiliation{School of Physics and Astronomy, University of
Nottingham, Nottingham NG7 2RD, United Kingdom}

\date{\today}

\begin{abstract}

The current noise spectrum of a single-electron transistor (SET)
coupled to a nanomechanical resonator is calculated in the
classical regime. Correlations between the charge on the SET
island and the position of the resonator give rise to a
distinctive noise spectrum which can be very different from that
of the uncoupled SET.  The current noise spectrum of the coupled
system contains peaks at both the frequency of the resonator and
double the resonator frequency, as well as a strong enhancement of
the noise at low frequencies. The heights of the peaks are
controlled by the strength of the coupling between the SET and the
resonator, the damping of the resonator, and the temperature of
the system.

\end{abstract}


\maketitle

\section{Introduction}
Nanoelectromechanical systems in which a nanomechanical resonator
is coupled to a mesoscopic conductor, such as a single-electron
transistor (SET), form a new class of mesoscopic system in which
there is a fascinating interplay between the electrical and
mechanical degrees of freedom. Coupling to nanomechanical degrees
of freedom can modify the transport properties of mesoscopic
conductors substantially, giving rise to a number of novel
phenomena such as electron shuttling\cite{shut,qs} and phonon
blockade effects.\cite{pb1,pb2} Furthermore, nanoelectromechanical
systems have important potential applications as ultra-sensitive
force detectors.\cite{elect}

A nanomechanical resonator coated with a thin metal layer can be
coupled to a SET as a mechanically compliant
voltage-gate.\cite{set,set2} Under such circumstances the
tunnelling rates of electrons through the SET island, and hence
the current flowing, depend sensitively on the position of the
resonator.\cite{set,set2,white,bw} It has been suggested that the
SET could act as a quantum-limited displacement detector for
micron-sized resonators\cite{ZB,set,set2} thus making it
potentially very important both for read-out of the resonator's
motion in force sensing applications\cite{bo} and for exploring
quantum effects in electromechanical systems.\cite{ABS,mpb}
However, fluctuations in the charge on the SET island exert a
stochastic force back on the resonator, affecting its motion, and
hence also the average current through the SET.\cite{ZB,ABZ}
Furthermore, the coupling between the resonator and the SET also
has an important influence on the SET current noise.

The effect on the dynamics of a nanomechanical resonator of
coupling to a mesoscopic conductor has been investigated by a
number of authors.\cite{noise,noise2,noise3,ABZ,hopkins,shut,qs}
For systems where the nanomechanical resonator forms a compliant
voltage gate adjacent to either a SET\cite{noise3,ABZ} or a
quantum point contact,\cite{noise} it has been found that the
interaction with the electrons drives the resonator into a
steady-state very similar to a thermal state, with an effective
temperature and damping constant. In particular, when the
resonator is treated as a quantum system it is found that whenever
the energy associated with the bias voltage applied across the
conductor is much greater than the energy quanta of the resonator,
the electrons rapidly dephase the resonator and heat it to an
effective temperature proportional to the bias
voltage.\cite{noise,noise2,noise3} An analysis of the dynamics of
a resonator coupled to a SET where the resonator is treated as a
classical system leads to very similar conclusions in the regime
where the applied bias voltage is relatively large and the charge
dynamics of the SET is adequately described by the orthodox
model.\cite{ABZ}

Although the dynamics of a nanomechanical resonator coupled to a
mesoscopic conductor has been investigated for a number of
different systems and regimes, there has been little systematic
study of the concomitant current-noise spectrum of the mesoscopic
conductor. Knowledge of the current-noise spectrum in
nanoelectromechanical systems is of interest both because it sets
limits on the sensitivity with which the current can be used as
measure of the motion of the resonator and, more generally,
because it has long been recognized that the current noise of
mesoscopic conductors provides important additional information,
beyond that available from simple average-current properties,
about the interactions which the conduction electrons
undergo.\cite{bandb} In this paper the current noise of a SET
coupled to a nanomechanical resonator is calculated in the
classical regime using an extension of the master equation
approach developed in Ref.\ \onlinecite{ABZ}. It is found that the
current noise spectrum of the SET-resonator system can be very
different from that of the uncoupled SET. In particular, the
interactions between the resonator and the transport electrons can
strongly enhance the SET current noise at low frequencies and give
rise to additional peaks in the noise spectrum at the resonator
frequency and at twice that value.

This paper is organized as follows. In Sec.\ II  the dynamical
model of the coupled resonator-SET system that forms the basis of
the following calculations is introduced. Then in Sec. III, the
relationship between the overall current noise spectrum of the SET
and the spectrum of the charge noise on the island and the noise
spectra of the tunnel currents between the island and the leads is
described. In Secs.\ IV and V the noise spectra of the tunnel
currents through the SET junctions and the charge noise on the SET
island are obtained, respectively. The full current noise spectrum
of the coupled SET-resonator system is obtained in Sec. VI.
Finally, Sec.\ VII contains the conclusions.

\section{Model of the coupled SET-resonator system}

\begin{figure}[t]
\center{ \epsfig{file=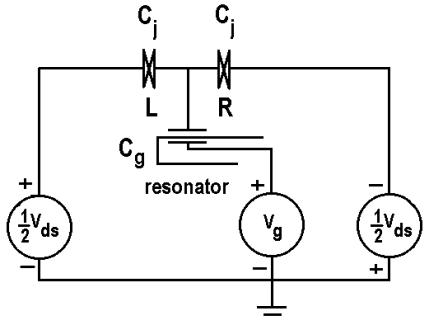, width=7.0cm}} \caption{Circuit
diagram for the coupled SET-resonator system. The SET island lies
between the two tunnel junctions which are assumed to have the
same capacitance $C_j$. Under the drain-source voltage shown,
$V_{\rm ds}$, the electrons tend to flow through the SET island
from right to left. The resonator is situated adjacent  to the SET
island and acts as a gate capacitor.} \label{fig:schema}
\end{figure}

A schematic diagram of the SET-resonator system is shown in Fig.\
\ref{fig:schema}. For simplicity, we have assumed a symmetric
arrangement of capacitances and voltages. The SET island is
connected to current leads by tunnel junctions with capacitance
$C_j$ and resistance $R$. The total voltage applied across the SET
island is $V_{\rm ds}$. The nanomechanical resonator forms the
gate capacitor, with a capacitance $C_g(x)$, and a voltage $V_g$
applied to it. The gate capacitance depends on the displacement of
the resonator, $x$, about its equilibrium separation from the SET
island, $d$. The resonator is displaced from its equilibrium
position by fluctuations in the SET charge and by thermal
fluctuations of the resonator itself, but for realistic
systems\cite{set,set2,ABZ} the resulting motion is on a scale much
less than the equilibrium separation (i.e. $x\ll d$), hence an
expansion of the capacitance to linear order is sufficient and we
may write $C_g(x)=C_g(1-x/d)$, where $C_g$ is a
constant.\cite{bw,ZB,ABZ} The gate capacitance is also assumed to
be much less than that of the junction capacitances, $C_g\ll C_j$,
which reflects the situation in typical devices.

 Under a relatively wide range of conditions, the coupled dynamics of a
nanomechanical resonator and a SET can be described using a master
equation formalism.\cite{ABZ}  Within this formalism, the SET is
assumed to be within the regime where the orthodox model is valid
and the resonator is treated as a classical harmonic oscillator,
with angular frequency $\omega_0$, and effective mass $m$. The
effects of damping processes on the resonator, beyond those due to
interactions with the SET, can be taken into account by including
an extrinsic damping rate, $\gamma_e$. The orthodox model
describes electron transport through the SET away from the Coulomb
blockade regions so that processes other than sequential
tunnelling of electrons may be neglected.\cite{FG} The
nanomechanical resonator is expected to behave classically
whenever the bias voltage applied to the SET is much greater than
the energy quanta of the resonator, i.e.\ $eV_{\rm ds}\gg \hbar
\omega_0$.\cite{noise} Complete details of the derivation,
justification and analysis of this classical model for the coupled
SET-resonator system are given in Ref.\ \onlinecite{ABZ}.

When the background temperature of the SET-resonator system,
$T_e$, is much less than the charging energy of the SET island,
$e^2/2C_j\gg k_{\rm B}T_e$, and the bias voltage is not too large,
the charge state of the SET island can be assumed to be limited to
two possible values, $N$ and $N+1$ excess electrons, and a pair of
coupled master equations can be derived for the SET-resonator
system. The bias voltage is also assumed to be much greater than
the background temperature, $eV_{\rm ds}\gg k_{\rm B}T_e$, and so
the effects of thermal fluctuations on the electron tunnel rates
are neglected.\cite{FG,ABZ} Writing $P_{N(N+1)}(x,u;t)$ as the
probability of having $N(N+1)$ electrons on the SET island and
finding the resonator at position $x$ and with velocity $u$, the
master equations for the evolution of the probability
distributions take the form\cite{ABZ}
\begin{eqnarray}
\frac{\partial P_N}{\partial t}&=&\omega_0^2x\frac{\partial
P_N}{\partial u}-u\frac{\partial P_N}{\partial
x}+\frac{\partial}{\partial u
}\left(\gamma_euP_{N}+\frac{G}{2m^2}\frac{\partial P_N }{\partial
u }\right)+\left(\Gamma_L
P_{N+1}-\Gamma_R P_N-\frac{m\omega_0^2 x_0}{Re^2} x P\right )\label{spn} \\
\frac{\partial P_{N+1}}{\partial
t}&=&\omega_0^2(x-x_0)\frac{\partial P_{N+1}}{\partial
u}-u\frac{\partial P_{N+1}}{\partial x}+\frac{\partial}{\partial u
}\left(\gamma_euP_{N+1}+\frac{G}{2m^2}\frac{\partial P_{N+1}
}{\partial u }\right)-\left(\Gamma_L P_{N+1}-\Gamma_R
P_N-\frac{m\omega_0^2 x_0}{Re^2} xP\right), \label{spn1}
\end{eqnarray}
where $P(x,u;t)=P_N(x,u;t)+P_{N+1}(x,u;t)$, $G=2mk_{\rm
B}T_e\gamma_e$ and $\Gamma_{L(R)}$ is the position-independent
part of the tunnel rate through the $L(R)$ junction. The
length-scale $x_0$ has a simple interpretation: it is the amount
by which the equilibrium separation between the SET and the
resonator changes when the charge on the SET island changes from
$N$ to $N+1$.

The master equations [Eqs.\ (\ref{spn}) and (\ref{spn1})] assume
that the bias voltage applied across the system is the dominant
energy-scale and the probability distributions for the resonator
are strongly peaked at $x\sim x_0$. Under such circumstances,
which are readily met in practicable resonator-SET
systems,\cite{ABZ} tunnelling processes that go against the
direction imposed by the bias voltage may be neglected.  The
resulting master equations can be used to extract equations of
motion for the average properties of the resonator and SET, and to
characterize the steady-state of the system. Furthermore, the
master equations provide the necessary information about the
coupled dynamics of the SET-resonator system to allow the
derivation of the SET current noise spectrum.

Analysis of the coupled master equations\cite{ABZ} shows that the
SET electrons act on the resonator like another thermal bath (in
addition to the substrate to which the resonator is attached)
which can be characterized by an intrinsic temperature, $T_i$,
proportional to $V_{\rm ds}$, and an intrinsic damping constant,
$\gamma_i$. The overall effective temperature of the resonator is
then determined by an average of the temperatures of the two baths
at temperatures $T_i$ and $T_e$, with weights that depend on the
respective damping constants $\gamma_i$ and $\gamma_e$.\cite{ABZ}

 The strength of
the interaction between the resonator and the SET is conveniently
described by the dimensionless parameter
$\kappa=m\omega_0^2x_0^2/(eV_{\rm ds})$ and the frequency by
$\epsilon=\omega_0 \tau_t$, where $\tau_t=Re/V_{\rm ds}$ is the
electron tunnelling time. In terms of these dimensionless
parameters, the intrinsic damping rate of the resonator due to
interactions with the electrons is \cite{ABZ} $\gamma_i=\kappa
\epsilon^2/\tau_t.$ The interaction of the resonator with the
electrons also leads to a slight renormalization  of resonator
frequency, $\epsilon'=\epsilon(1-\kappa)^{1/2}$. Finally, the
external temperature is conveniently expressed in terms of the
drain-source voltage as the dimensionless parameter,
$\Theta=k_{\rm B}T_e/eV_{\rm ds}$.

\section{SET current noise}
The noise in the current through a SET is generated by the
fluctuations in the time dependent current flowing through the
system. Since the SET contains more than one tunnel junction, the
time-dependent current which flows is not simply the tunnel
current across either one of the
junctions.\cite{reallyuseful,bandb,korot,sun} When an electron
tunnels through either junction, charge also flows in the external
circuit connected to the SET island (so that the charge stored in
each capacitor returns to its equilibrium value) and hence also
contributes to the current. The current flowing through the SET
can be measured in either the left or right-hand lead and the
average current flowing through the leads must necessarily be the
same.

The time dependent current through the gated SET  is, in
principle, slightly different in the left and right-hand leads,
even for the symmetrical case we consider here where the
tunnel-junction capacitances, $C_j$, are the
same.\cite{reallyuseful} The time dependent current through the
left-hand lead can be written\cite{reallyuseful,bandb}
\begin{equation}
I^{(L)}(t)=a^{(L)}\dot{Q}_L(t)+b^{(L)}\dot{Q}_R(t), \label{il}
\end{equation}
whilst the time dependent current through the right-hand lead is
given by\cite{reallyuseful}
\begin{equation}
I^{(R)}(t)=a^{(R)}\dot{Q}_L(t)+b^{(R)}\dot{Q}_R(t),  \label{ir}
\end{equation}
where $\dot{Q}_{L(R)}$ are the tunnel currents through the $L(R)$
junctions respectively, and\cite{reallyuseful}
\begin{eqnarray}
a^{(L)}=b^{(R)}&=&\frac{C_j+C_g(x)}{2C_j+C_g(x)} \\
b^{(L)}=a^{(R)}&=&\frac{C_j}{2C_j+C_g(x)}.
\end{eqnarray}
Notice, however, that the two coefficients $a^{(i)}$ and
$b^{(i)}$, $i=L,R$, add up to unity, whichever lead we consider.
Because of the dependence of these coefficients on the gate
capacitance, $C_g(x)$, there will necessarily be a weak dependence
of the coefficients on the position of the resonator. However,
even the leading order position dependent term in the coefficients
will be smaller by a factor $C_g/C_j$ than those arising within
the expressions for the time dependent currents. Since we are
treating the case where $C_g/C_j\ll 1$, the corrections generated
by the position dependence of $a^{(i)}$ and $b^{(i)}$, $i=L,R$,
will be neglected here.\cite{neglect} In fact, it turns out that
the noise spectrum is relatively insensitive to the exact values
of the coefficients, and hence which lead the current is measured
in. Therefore, for simplicity, we will not distinguish between the
two leads in what follows and set $a^{(L)}=a^{(R)}=a$, and
$b^{(L)}=b^{(R)}=b$.

The current fluctuations in the SET are constrained by the
conservation of charge. Charge conservation implies that the
charge that has tunnelled through the left-hand junction in time
$t$, $Q_L(t)$, and the charge that has tunnelled through the right
junction in the same period, $Q_R(t)$ must be connected by the
relation
\begin{equation}
Q_L(t)-Q_R(t)=Q_D(t)
\end{equation}
where $Q_D(t)$ is the charge dwelling on the island between the
two barriers at time $t$. Furthermore, since the average tunnel
currents through the left- and right-hand junctions are the same,
they are both also equal to the average current through the
SET,\cite{ABZ}
\begin{equation}
\langle I \rangle=\langle \dot{Q}_L\rangle=\langle
\dot{Q}_R\rangle,
\end{equation}
and hence we also have, $\langle \dot{Q}_D\rangle=0$.

The spectrum of current noise in the SET is given by the Fourier
transform of the current-current correlation function,\cite{korot}
\begin{equation}
S_{II}(\omega)=2\int_{-\infty}^{\infty}d\tau
K_{II}(\tau)\cos(\omega \tau),
\end{equation}
where
\begin{equation}
K_{II}(\tau)=\langle I(t+\tau)I(t)\rangle-\langle I(t) \rangle^2.
\end{equation}
The current-current correlation function, $K_{II}(\tau)$, is taken
to be independent of $t$.

The noise in the total current is a combination of the noise in
the tunnel currents through the left and right-hand junctions,
$S_{I_LI_L}(\omega)$ and $S_{I_RI_R}(\omega)$ respectively, and
the charge noise of the SET island $S_{Q}(\omega)$. Using the
expression for the full current in terms of tunnel currents
through the two junctions and the charge conservation relation, we
can write the current noise spectrum as a sum of the three
terms:\cite{brandes}
\begin{equation}
S_{II}(\omega)=aS_{I_LI_L}(\omega)+bS_{I_RI_R}(\omega)-ab\omega^2S_Q(\omega),\label{abeq}
\end{equation}
where
\begin{eqnarray}
S_{I_LI_L}(\omega)&=& 2\int_{-\infty}^{\infty}d\tau \cos(\omega
\tau)\left[\langle \dot{Q}_L(t+\tau)\dot{Q}_L(t)\rangle-\langle
\dot{Q}_L(t) \rangle^2\right] \\
S_{I_RI_R}(\omega)&=&
2\int_{-\infty}^{\infty}d\tau \cos(\omega \tau)\left[\langle
\dot{Q}_R(t+\tau)\dot{Q}_R(t)\rangle-\langle
\dot{Q}_R (t)\rangle^2\right]\\
\omega^{2}S_{Q}(\omega)&=&2\int_{-\infty}^{\infty}d\tau
\cos(\omega \tau)\langle \dot{Q}_D(t+\tau)\dot{Q}_D(t)\rangle.
\label{sqeqn}
\end{eqnarray}
In the last equation, the SET island charge noise, $S_Q(\omega)$,
is identified with the noise in $\dot{Q}_D(t)$ (this connection
will be derived in Sec.\ V). Notice that the charge noise acts to
reduce the full noise spectrum, compared to the individual tunnel
current noise spectra, reflecting the well-known fact that current
noise is suppressed in double barrier structures.\cite{bandb}

The zero-frequency noise of the tunnel currents through the two
junctions are equal, because of charge conservation, and hence
$S_{II}(0)=S_{I_LI_L}(0)=S_{I_RI_R}(0)$. However, in order to
obtain the full current noise at finite frequency we must
calculate each of its three components. The noise in the tunnel
currents $S_{I_LI_L}(\omega)$ and $S_{I_RI_R}(\omega)$ are
obtained first, using an equation of motion method, then the
charge noise, $S_{Q}(\omega)$ is calculated using a similar
approach. The three elements of the current noise are then
combined to obtain the full spectrum.

\section{Tunnel current noise spectra}
The noise in the tunnel current across the SET junctions can be
calculated using an approach due to MacDonald\cite{mac,macp} which
relies on the fact that the number of charges that have passed
across the junction as a function of time can be counted. For the
SET-resonator system, the description in terms of a probability
distribution is easily extended to record the number of electrons
passing across one or other of the tunnel barriers linking the SET
island to the leads.

Starting with the left-hand junction, we follow
MacDonald\cite{mac,macp} in introducing the statistically
stationary variable,
\begin{equation}
\dot{\tilde{Q}}_L=\dot{Q}_L-\langle \dot{Q}_L\rangle
\end{equation}
and write,
\begin{equation}
S_{I_LI_L}(\omega)=2\omega\int_0^{\infty}\frac{\partial\langle
\tilde{Q}_L^2(\tau)\rangle}{\partial \tau}\sin(\omega \tau)d\tau
\label{noisespec}
\end{equation}
where,
\begin{eqnarray}
\langle
\tilde{Q}_L^2(\tau)\rangle&=&\left\langle\left[\int_t^{t+\tau}\dot{\tilde{Q}}_L(t')dt'\right]^2\right\rangle\\
&=&\left\langle \left [ \int_t^{t+\tau}
\dot{Q}_L(t')dt'-\langle \dot{Q}_L\rangle\tau\right]^2\right\rangle\\
&=&e^2\langle n^2(\tau)\rangle -I_0^2\tau^2
\end{eqnarray}
with $n(\tau)$ the number of electrons which passed across the
left-hand junction in time $\tau$ and $I_0=\langle
I\rangle=\langle \dot{Q}_L\rangle$, the average current. The
MacDonald approach has been used to calculate the current-noise in
a number of mesoscopic systems\cite{mac1,mac2} and the subtleties
of the method are described particularly carefully by Ruskov and
Korotkov.\cite{mac2}

The probability distributions for the SET-resonator system are
readily generalized to include the number of electrons that have
passed through either one of the junctions. In order to calculate
the noise in the current through the left-hand junction, we define
$P^n_{N(N+1)}(x,u;t)$ as the probability that, after time $t$, $n$
electrons have passed across the left-hand junction, there are
$N(N+1)$ electrons on the SET island and the resonator is at
position $x$ with velocity $u$. The master equations [Eqs.\
(\ref{spn}) and (\ref{spn1})] now take the modified form
\begin{eqnarray}
\frac{\partial P^n_N}{\partial t}&=&\omega_0^2x\frac{\partial
P^n_N}{\partial u}-u\frac{\partial P^n_N}{\partial
x}+\frac{\partial}{\partial u
}\left(\gamma_euP_{N}^n+\frac{G}{2m^2}\frac{\partial P_N^n
}{\partial u }\right)+\left(\Gamma_L P^{n-1}_{N+1}-\Gamma_R P_N^n-
\frac{m\omega_0^2 x_0}{Re^2} x [P^{n-1}_{N+1}+P^{n}_{N}]\right )\label{spnna} \\
\frac{\partial P^n_{N+1}}{\partial
t}&=&\omega_0^2(x-x_0)\frac{\partial P^n_{N+1}}{\partial
u}-u\frac{\partial P^n_{N+1}}{\partial x}+\frac{\partial}{\partial
u }\left(\gamma_euP_{N+1}^n+\frac{G}{2m^2}\frac{\partial P_{N+1}^n
}{\partial u }\right)-\left(\Gamma_L P^n_{N+1}-\Gamma_R
P^n_N-\frac{m\omega_0^2 x_0}{Re^2} xP^n\right), \label{p1}
\end{eqnarray}
where $P^n(x,u;t)=P^n_{N}(x,u;t)+P^n_{N+1}(x,u;t)$. Notice that
only the terms which describe the transition from charge state
$N+1$ to $N$ with an electron tunnelling out through the left
junction couple to a change in $n$.

 The rate of change of the
average of the squared number of charges passing through the
left-hand junction in time $\tau$ is given by\cite{mac2}
\begin{eqnarray}
\frac{d\langle n^2(\tau)\rangle}{d\tau}&=&\sum_n n^2\int dx \int
du
\left [ \dot{P}^{n}_{N+1}(x,u;\tau)+\dot{P}^{n}_{N}(x,u;\tau)\right ] \\
&=& 2\left(\Gamma_L\langle
n\rangle_{N+1}-\frac{m\omega_0^2x_0}{Re^2}\langle
xn\rangle_{N+1}\right)+\frac{I_0}{e}, \label{nxn}
\end{eqnarray}
where we have re-ordered the sum over $n$ in the first term, and
substituted from Eqs.\ (\ref{spnna})  and (\ref{p1}). The averages
$\langle \dots \rangle_{N+1}$, are with respect to the
sub-ensemble of systems with $N+1$ charges on the SET island,
i.e.\
\[
\langle \dots \rangle_{N+1}=\sum_n\int dx\int du (\dots
)P^n_{N+1}(x,u;\tau).
\]
The initial conditions for the distributions,
$P^n_{N(N+1)}(x,u;\tau=0)$, are given by the corresponding
steady-state distributions for the SET-resonator system,\cite{ABZ}
$P^{(s)}_{N(N+1)}(x,u)$,
\[
P^n_{N(N+1)}(x,u;\tau=0)=\delta_{n,0}P^{(s)}_{N(N+1)}(x,u),
\]
and hence we have
\[
\sum_nP^n_{N(N+1)}(x,u;\tau)=P^{(s)}_{N(N+1)}(x,u),
\]
for all $\tau$.

 The overall expression for the noise in the current through
the left-hand junction is given by
\begin{equation}
S_{I_LI_L}(\omega)=2\omega\int_0^{\infty}\left\{2\left[e^2\left(\Gamma_L\langle
n\rangle_{N+1}-\frac{\kappa}{\tau_t}\frac{\langle
xn\rangle_{N+1}}{x_0}\right)-I_0^2\tau
\right]+eI_0\right\}\sin(\omega \tau)d\tau.  \label{noiselh}
\end{equation}
The quantities $\langle n(\tau)\rangle_{N+1}$ and $\langle
xn(\tau)\rangle_{N+1}$ can be obtained by deriving their equations
of motion and those of the other average quantities they couple
to. The equations of motion for moments of the form $\langle
x^pu^q n\rangle$ and $\langle x^pu^q n\rangle_{N+1}$ are given by
\begin{eqnarray}
\frac{d}{d\tau}\langle {x^pu^qn}\rangle&=&\sum_n\int dx\int du
x^pu^qn
\left[\dot{P}^n_{N+1}(x,u;\tau)+\dot{P}^n_{N}(x,u;\tau)\right],\\
\frac{d}{d\tau}\langle {x^pu^qn}\rangle_{N+1}&=&\sum_n\int dx\int
du x^pu^qn \dot{P}^n_{N+1}(x,u;\tau)
\end{eqnarray}
respectively. Substituting for $\dot{P}^n_{N+1(N)}(x,u;\tau)$ from
the appropriate master equation [Eqns. (\ref{spnna}) and
(\ref{p1})], we can derive the closed set of equations,
\begin{eqnarray}
\frac{d}{d\tau}\langle
{n}\rangle_{N+1}&=&\frac{I_0}{e}\Gamma_R\tau
-(\Gamma_L+\Gamma_R)\langle n\rangle_{N+1}+
\frac{m\omega_0^2x_0}{Re^2}\langle xn\rangle
\label{one} \\
\frac{d}{d\tau}\langle {xn}\rangle &=& \langle un \rangle
+\Gamma_L\langle x\rangle_{N+1}-\frac{m\omega_0^2x_0}{Re^2}\langle
x^2
\rangle_{N+1} \label{two}\\
\frac{d}{d\tau}\langle {un}\rangle&=&-\omega_0^2 [\langle
xn\rangle -x_0\langle n\rangle_{N+1}]
-\frac{m\omega_0^2x_0}{Re^2}\langle
ux\rangle_{N+1}+\Gamma_L\langle u\rangle_{N+1} -\gamma_e\langle un\rangle \label{three}\\
\frac{d}{d\tau}\langle {xn}\rangle_{N+1}&=&\langle
un\rangle_{N+1}+\Gamma_R\langle xn\rangle
-(\Gamma_L+\Gamma_R)\langle
xn\rangle_{N+1}+\frac{m\omega_0^2x_0}{Re^2}\langle x^2n\rangle\\
\frac{d}{d\tau}\langle
{un}\rangle_{N+1}&=&-\omega_0^2\left[\langle xn\rangle_{N+1}
-x_0\langle n\rangle_{N+1}\right] +\Gamma_R\langle u n\rangle
+\frac{m\omega_0^2x_0}{Re^2}\langle xun\rangle
-(\Gamma_L+\Gamma_R)\langle un\rangle_{N+1} -\gamma_e\langle
{un}\rangle_{N+1}\\
\frac{d}{d\tau}\langle {x^2n}\rangle&=& 2\langle
xun\rangle+\Gamma_L\langle x^2\rangle_{N+1}
-\frac{m\omega_0^2x_0}{Re^2}\langle x^3\rangle_{N+1} \\
\frac{d}{d\tau}\langle {u^2n}\rangle&=& -2\omega_0^2[\langle
xun\rangle-x_0\langle un\rangle_{N+1}]+\Gamma_L \langle
 u^2\rangle_{N+1} -\frac{m\omega_0^2x_0}{Re^2}
 \langle u^2x\rangle_{N+1} -2\gamma_e\langle {u^2n}\rangle+\frac{I_0G}{em^2}\tau\\
\frac{d}{d\tau}\langle {xun}\rangle&=&-\omega_0^2[\langle
x^2n\rangle-x_0\langle nx\rangle_{N+1}] +\langle
u^2n\rangle+\Gamma_L\langle
xu\rangle_{N+1}-\frac{m\omega_0^2x_0}{Re^2}\langle
x^2u\rangle_{N+1}-\gamma_e\langle {xun}\rangle. \label{final}
\end{eqnarray}
The averages which do not involve $n$ are constants (i.e.
independent of $\tau$) and are evaluated for the steady-state
distributions, $P^{(s)}_{N(N+1)}(x,u)$.\cite{mac2} The details of
the evaluation of these averages are given in Appendix A.

The set of coupled differential equations [Eqs.\
(\ref{one}-\ref{final})] is readily integrated numerically to give
$\langle n \rangle_{N+1}$ and $\langle xn\rangle_{N+1}$, from
which the left-hand junction current noise spectrum can be
obtained.  Since we are interested in the number of electrons
passing through the left-hand junction in a time $\tau$, all
quantities involving $n$ will clearly be set to zero at
$\tau=0$.\cite{mac2}

\begin{figure}[t]
\center{\epsfig{file=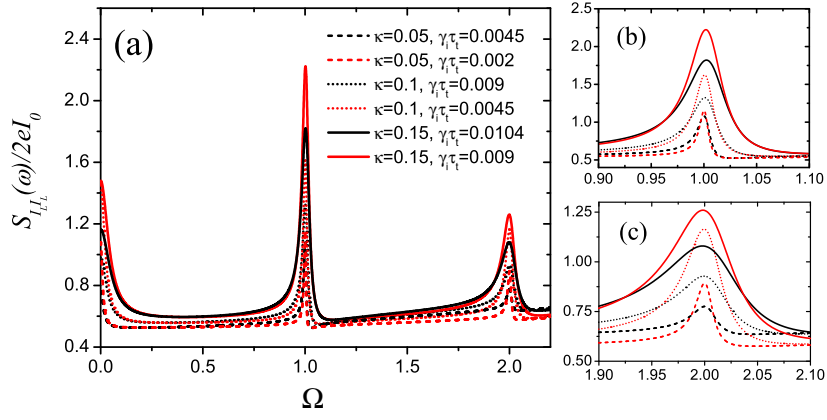,width=12.0cm} }
  \caption{Current noise spectrum of the left-hand junction, $S_{I_LI_L}(\omega)$,
  for a range of resonator parameters as a function of the scaled frequency
  $\Omega=\omega\tau_t/\epsilon'$. The full spectrum is shown in
  (a) while the insets (b) and (c) show the details of the peaks at $\Omega=1$
  and 2, respectively. All calculations are performed for $\gamma_e\tau_t=\Theta=0$ and
$\Gamma_L=\Gamma_R$.} \label{fig:cnoise1}
\end{figure}
Integrating the equations of motion for $\langle n\rangle_{N+1}$
and $\langle xn\rangle_{N+1}$ and performing the Fourier integral
numerically\cite{nr} (incorporating an appropriate long-time
cut-off), we can obtain the current-noise spectrum,
$S_{I_LI_L}(\omega)$ for various choices of the system parameters.
The current noise with the coupling to the resonator set to zero,
$S^0_{I_LI_L}(\omega)$, can also be calculated analytically for
comparison,\cite{korot}
\begin{eqnarray}
S^0_{I_LI_L}(\omega)&=&2\omega I_0\int_0^\infty d\tau\sin(\omega
\tau)\left [e+\frac{2I_0}{\Gamma_L+\Gamma_R}\left({\rm
e}^{-(\Gamma_L+\Gamma_R)\tau}-1\right)\right]\\
&=&2eI_0-\frac{4I_0^2(\Gamma_L+\Gamma_R)}{(\Gamma_L+\Gamma_R)^2+\omega^2}.
\end{eqnarray}

Initially, we consider the simplest case, where there is no
external damping and the resonator substrate is at zero
temperature (i.e.\ $\gamma_e\tau_t=0$ and $\Theta=0$), and examine
the dependence of the tunnel current noise spectrum on the
resonator frequency and the strength of the coupling to the SET.
Figure \ref{fig:cnoise1} shows $S_{I_LI_L}(\omega)$  for several
choices of the dimensionless coupling and frequency parameters,
$\kappa$ and $\epsilon$.\cite{pchoice}  For ease of comparison,
the frequency axis is scaled by the renormalized resonator
frequency, $\Omega=\omega\tau_t/\epsilon'$. The spectra are
dominated by sharp, asymmetric, peaks arising at $\Omega=0$, the
renormalized resonator frequency $\Omega=1$ and at twice that
frequency, $\Omega=2$. None of these three features appear in the
current noise spectrum for a SET without a resonator. At high
frequencies, beyond the range shown in Fig.\ \ref{fig:cnoise1},
all of the spectra reduce to the usual single-junction value
$2eI_0$.

The effect of the resonator on the noise spectrum depends on both
the strength of the coupling between the resonator position and
the SET junction tunnel rates, measured by $\kappa$, as well as
the underlying dynamics of the resonator, which in the absence of
extrinsic temperature or damping, is conveniently parameterized by
$\gamma_i\tau_t=\kappa\epsilon^2$. The coupling strength,
$\kappa$, plays a dual role as it controls both the amount by
which the resonator motion alters the electron tunnel rates, and
the resonator dynamics through the intrinsic damping $\gamma_i$.
Fig.\ \ref{fig:cnoise1} shows that the heights of the peaks depend
sensitively on both $\kappa$ and $\gamma_i$. The peak heights
increase with an increase in $\kappa$, at fixed $\gamma_i$, whilst
an increase in $\gamma_i$ at fixed $\kappa$ (i.e.\ an increase in
$\epsilon$) reduces the heights of the peaks. Notice, however,
that the peaks at $\Omega=0$ and 2 reduce more strongly as
$\gamma_i$ increases than the one at $\Omega=1$.

\begin{figure}[t]
\center{\epsfig{file=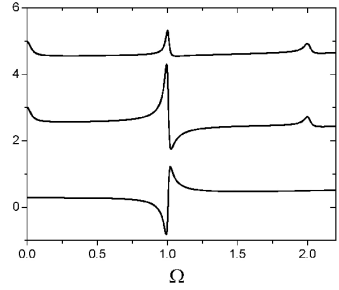, width=6.0cm} } \caption{Comparison
of the contributions to the left-hand junction current noise due
to the $\langle xn \rangle_{N+1}$ and $\langle n\rangle_{N+1}$
terms in Eq.\ (\ref{nxn}), given by $F(\omega)$ and $G(\omega)$
respectively. The curves have been displaced vertically for
clarity. From top to bottom, the curves are:
$S_{I_LI_L}(\omega)/(2eI_0)$, $F(\omega)$ and $G(\omega)$. In each
case  $\kappa=0.1$, $\epsilon=0.3$, $\gamma_e\tau_t=\Theta=0$ and
$\Gamma_L=\Gamma_R$.} \label{fig:mean}
\end{figure}

We can understand the origin of the features in the current noise
spectrum by splitting it into its two constituent parts: that
arising from $\langle n\rangle_{N+1}$ and that from $\langle xn
\rangle_{N+1}$. Figure \ref{fig:mean} shows a noise spectrum for a
particular choice of parameters ($\kappa=0.1, \epsilon=0.3$), and
the two functions $F(\omega)$ and $G(\omega)$, which add to give
the noise spectrum
\begin{equation}
\frac{S_{I_LI_L}(\omega)}{2eI_0}=F(\omega)+G(\omega).
\end{equation}
The functions $F(\omega)$ and $G(\omega)$ are essentially the
Fourier integrals of $\langle xn \rangle_{N+1}$ and $\langle n
\rangle _{N+1}$, respectively, with the appropriate linear and
constant terms subtracted off in each case. Written out
explicitly, they have the form
\begin{eqnarray}
F(\omega)&=&\frac{2\omega}{I_0}\int_0^{\infty}d\tau\left[\frac{\kappa}{2}
I_0-\frac{\kappa}{\tau_t}\left(
e\frac{\langle xn\rangle_{N+1}}{x_0}-\alpha\tau\right) \right]\sin(\omega\tau)\\
G(\omega)&=&
\frac{2\omega}{I_0}\int_0^{\infty}d\tau\left[\frac{(1-\kappa)}{2}
I_0+\Gamma_L\left( e\langle n\rangle_{N+1}-\alpha\tau\right)
\right]\sin(\omega\tau),
\end{eqnarray}
with
\begin{equation}
\alpha=\frac{I_0^2}{e\left(\Gamma_L-\kappa/\tau_t\right)}.
\end{equation}

The function $G(\omega)$, which contains the $\langle n
\rangle_{N+1}$ term, has a single feature, at the renormalized
resonator frequency. In contrast, the function $F(\omega)$, which
contains the $\langle xn\rangle_{N+1}$ term, has three features:
one at $\Omega=0$, one at the renormalized frequency, $\Omega=1$,
and one at $\Omega=2$. We can understand these features in terms
of non-trivial correlations between the variables $x$ and $n$.
From the spectrum $G(\omega)$, we see that the $n$ term has a
component oscillating at frequency $\epsilon'$ and since $x$ will
also oscillate at the same frequency, it is not surprising that
the combination of the two dynamical variables in the term
$\langle xn\rangle_{N+1}$ gives rise to a component oscillating at
$2\epsilon'$ as well as at $\epsilon'$. The explicit dependence on
$x$ in $\langle xn \rangle_{N+1}$ also makes the features in
$F(\omega)$ more sensitive to the magnitude of $\gamma_i$ than
those in $G(\omega)$. Hence in the full spectrum the peaks that
are due to $\langle xn \rangle_{N+1}$ alone (at $\Omega=0,2$)
reduce more rapidly with increasing $\gamma_i$ than the one with
contributions from $\langle xn \rangle_{N+1}$ and $\langle
n\rangle_{N+1}$ (at $\Omega=1$).

 The enhancement of the low
frequency noise is to be expected because of the effective
correlations between successive electrons induced by the
interactions with the resonator. It is known from other
contexts\cite{loss} that zero-frequency noise can be increased
when an electron tunnelling event is able to change the internal
state of a mesoscopic system (such as a quantum dot). In this
case, an electron passing through the SET alters the state of the
resonator, which in turn alters the tunnelling rates for
subsequent electrons. However, any excitation of the resonator is
damped out over time and hence the zero-frequency peak in the
spectrum follows a very similar pattern to the other peaks: it
increases with the coupling strength, $\kappa$, but reduces with
increasing damping, $\gamma_i$.

It is the approximation of the gate capacitance by the linear
form, $C_g(x)\simeq C_g(1-x/d)$, which means that there are no
further peaks in the current noise beyond that at $\Omega=2$. If
higher order terms had been included in the expression for the
gate capacitance then they would have lead to higher order terms
in the position dependence of the tunnel rates, which we would
expect to generate peaks at higher multiples of the resonator
frequency. However, within the regime we consider here, where
$x/d\ll 1$, the additional peaks at higher frequency would be much
smaller than those at $\Omega=1$ and $2$.

\begin{figure}[t]
\center{\epsfig{file=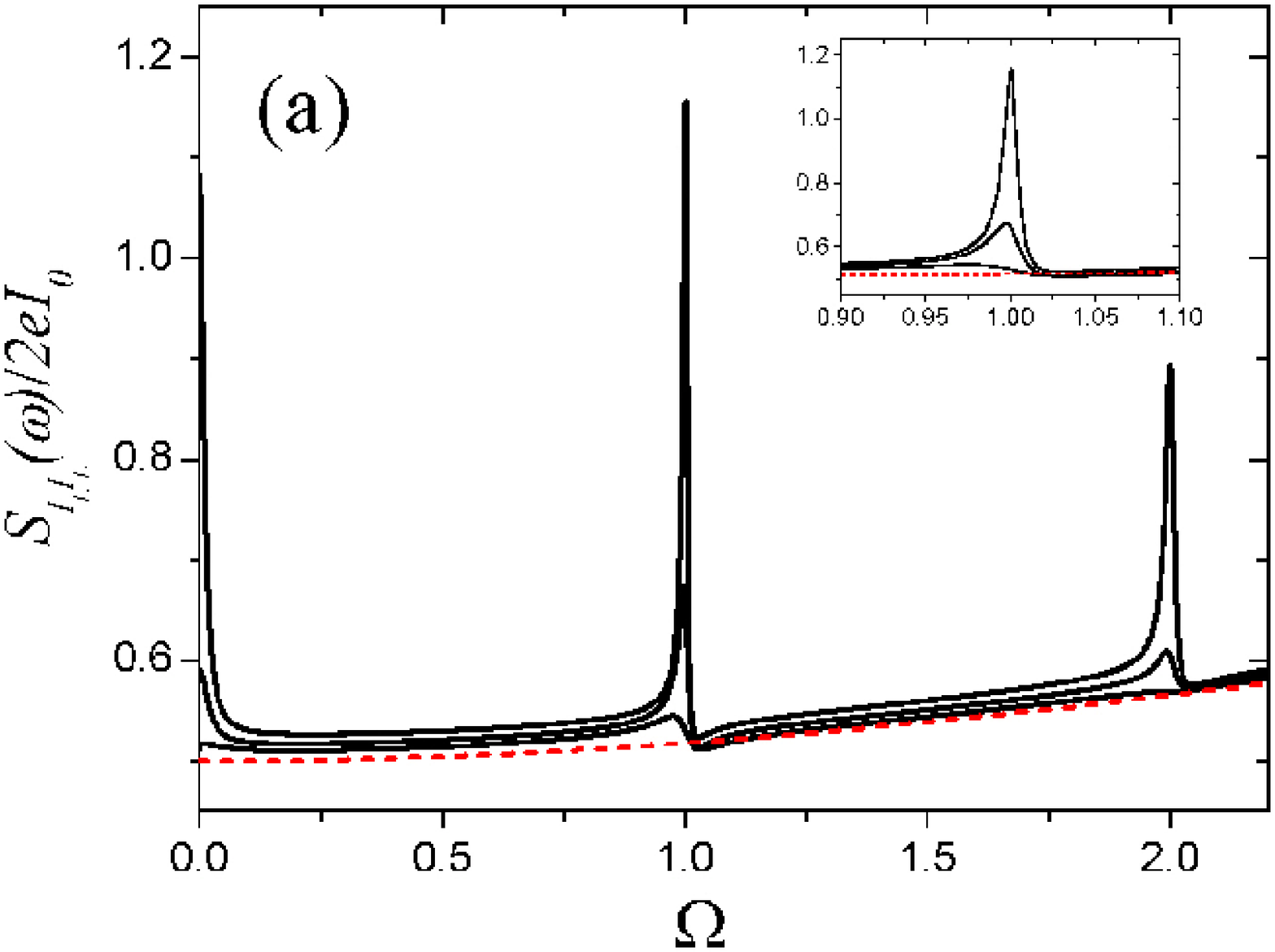,width=7.0cm}
\epsfig{file=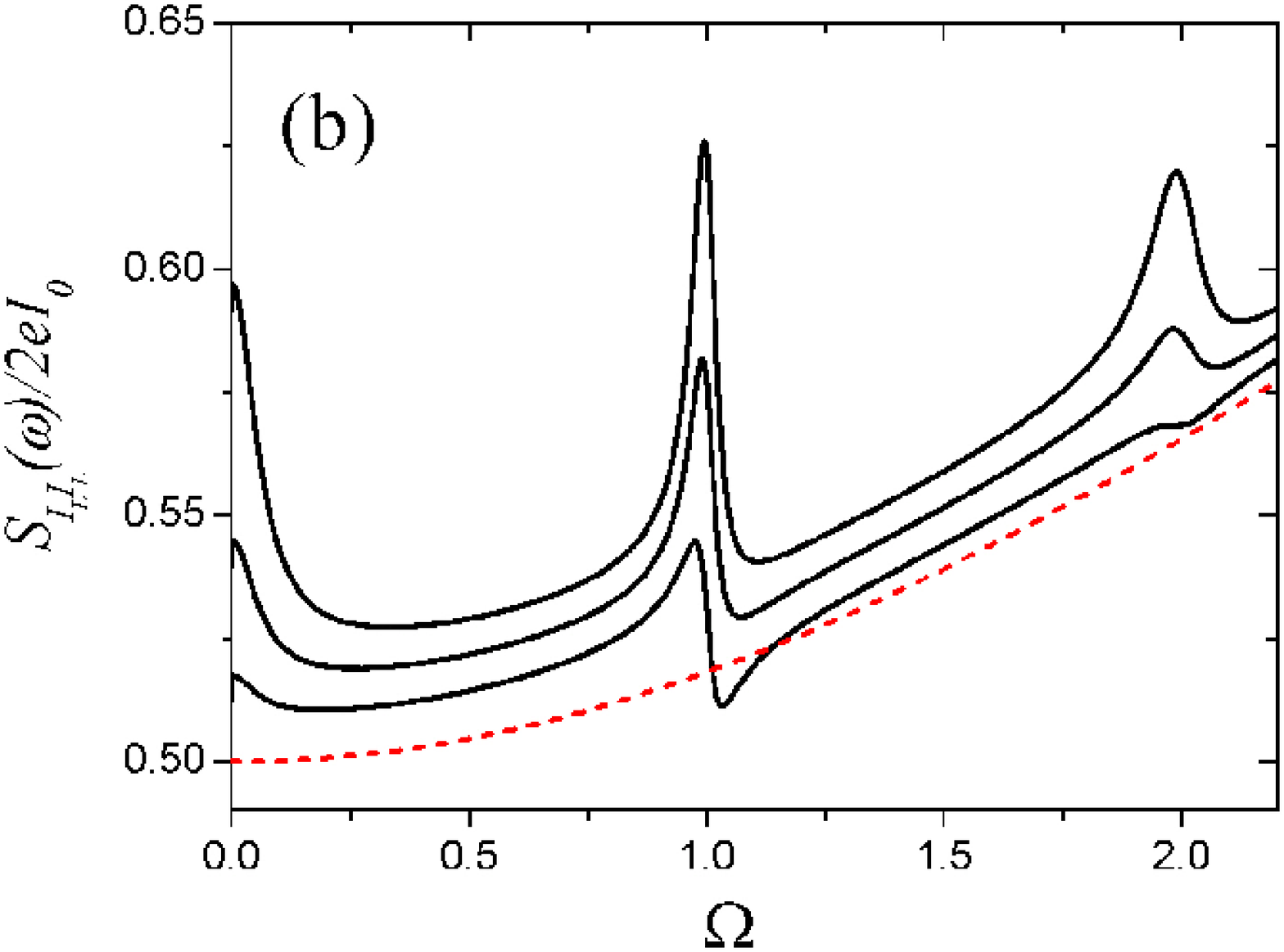,width=7.0cm}}
  \caption{Effects of finite extrinsic damping and temperature on the current noise through the left-hand junction,
   $S_{I_LI_L}(\omega)$,
  for the coupled system with $\kappa=0.05,\epsilon=0.2$. Panel
  (a) shows shows the noise spectra for $\gamma_e=0$ (upper curve), $\gamma_e=\gamma_i$
  (middle curve) and $\gamma_e=5\gamma_i$
  (lower curve), with $\Theta=0$ in each case. Panel (b) shows the noise
spectra for $\Theta=0.2$ (upper curve),  $\Theta=0.1$ (middle
curve) and $\Theta=0$ (lower curve) with $\gamma_e=5\gamma_i$ in
each case. In both plots the corresponding noise spectrum for the
uncoupled system is shown as a dashed curve.
    All calculations are performed for $\Gamma_L=\Gamma_R$.}
\label{fig:cndamp}
\end{figure}

The effects of extrinsic damping and finite background temperature
on the tunnel current noise spectrum are illustrated in Fig.\
\ref{fig:cndamp}, with the appropriate noise spectrum for the
uncoupled system shown for comparison. It is clear that a non-zero
value of $\gamma_e$ rounds down each of the peaks at $\Omega=0,1$
and 2. Furthermore, the peaks at $\Omega=1$ and $2$ are also
shifted to slightly lower frequencies, reflecting the fact that
the extrinsic damping causes an extra reduction in the resonator
frequency (beyond that due to the interaction with the electrons).
The influence of extrinsic damping gradually fades at higher
frequencies where the spectrum is hardly affected by the resonator
anyway. A non-zero background temperature counteracts the damping,
increasing the height and broadening the peaks.

The comparison with the uncoupled case in Fig.\ \ref{fig:cndamp},
also reveals a small suppression of the noise just after the peak
at $\Omega=1$, and a similar feature after the peak at $\Omega=2$.
We can understand the presence of this suppression in a loose way
by considering the actual motion of the resonator as a competition
between two components. Firstly, we can consider a resonator
coupled to a thermal bath, in this case the mechanical motion
would fluctuate strongly leading to an enhancement of the noise at
the mechanical frequency.\cite{hopkins} Of course these
thermal-like fluctuations arise already from the interaction
between the resonator and the electrons,\cite{ABZ} and are
increased by the presence of an extrinsic bath. However, it seems
likely that the interaction between the electrons and the
resonator must also induce some coordination between the
electronic and mechanical motion which would tend to make the
current flow more regular and hence suppress the noise, at least
at the resonator frequency. As we would expect from this simple
picture, Fig.\ \ref{fig:cndamp}b shows that increasing the
background temperature rapidly washes out the regions of noise
suppression.

The noise in the current through the right-hand junction can be
obtained in exactly the same manner as that through the left-hand
junction. Using MacDonald's formula, we have
\begin{equation}
S_{I_RI_R}(\omega)=2\omega\int_0^{\infty}\left\{2\left[e^2\left(\Gamma_R\langle
n\rangle_N+\frac{\kappa}{\tau_t}\frac{\langle
xn\rangle_N}{x_0}\right)-I_0^2\tau \right]+eI_0\right\}\sin(\omega
\tau)d\tau \label{noiserh},
\end{equation}
where in this case $n(\tau)$ is the number of electrons to pass
through the {\it right-hand} junction in time $\tau$.

 The master equations
[Eqs.\ (\ref{spn}) and (\ref{spn1})] are also readily modified to
include the number of electrons tunnelling (from right to left)
across the right-hand junction. A closed set of equations of
motion that allow  $\langle n\rangle_N$ and $\langle xn\rangle_N$
to be determined by numerical integration are obtained following
the same procedure as that used for the left-hand junction.

Carrying out the numerical integration and then the Fourier
integral, as before, we can compare the resulting noise spectrum
through the right-hand junction with that through the left-hand
junction. As would be expected for the very symmetrical SET layout
we consider here, the current noise spectra of the two junctions
are very similar, but away from $\Omega=0$ they are not identical.
In fact, the magnitudes of the peaks in the left and right-hand
tunnel noise spectra at $\Omega=1,2$ differ by up to a few percent
for the parameters used here. The differences in the two spectra
arise because the resonator breaks the overall symmetry of the
circuit.

\section{Charge noise on the SET island}
Apart from the contributions due to the tunnel currents across the
two SET junctions, the full current-noise spectrum also contains a
contribution arising from fluctuations in the charge stored on the
SET island,
\begin{equation}
\omega^2S_{Q}(\omega)=2\int_{-\infty}^{\infty}d\tau \cos(\omega
\tau)\langle \dot{Q}_D(t+\tau)\dot{Q}_D(t)\rangle.
\end{equation}
The average charge on the SET island is a constant (i.e.\ $\langle
\dot{Q}_D\rangle=0$), thus we can use MacDonald's formula to show
explicitly that this term is indeed the charge noise spectrum,
\begin{eqnarray}
S_{Q}(\omega)&=&2\omega^{-1}\int_{0}^{\infty}\frac{\partial\langle
Q_D(\tau)^2\rangle}{\partial \tau}  \sin(\omega
\tau)d\tau\\
&=&4\int_{0}^{\infty}d\tau \cos(\omega \tau)\left[\langle
Q_D(t+\tau)Q_D(t)\rangle-\langle Q_D^2\rangle\right],
\end{eqnarray}
where an integration by parts has been performed.

Within our model of the SET, the excess charge can only be either
zero or one electron, hence $\langle Q_D^2\rangle=-e\langle
Q_{D}\rangle$. Since $\langle Q_D(t)\rangle$ is simply the average
occupancy of the SET island multiplied by the electronic charge,
$-e$, we can use the master equations to write down a closed set
of coupled equations that link the average excess charge,
position, $\langle x\rangle$, and velocity, $\langle u\rangle$, of
the resonator,
\begin{eqnarray}
\frac{d}{dt}\langle Q_D(t)\rangle&=&-e\Gamma_R-\frac{V_{\rm
ds}}{eR}\langle
Q_D(t)\rangle-\frac{m\omega_0^2x_0}{eR}\langle x(t)\rangle \label{mf1} \\
\frac{d}{dt}\langle x(t)\rangle&=&\langle u(t)\rangle \label{mf2} \\
\frac{d}{dt}\langle u(t)\rangle&=&-\omega_0^2\langle x(t)\rangle
-\frac{\omega_0^2x_0}{e}\langle Q_D(t)\rangle -\gamma_e\langle
u(t)\rangle. \label{mf3}
\end{eqnarray}
Note that these equations are equivalent to the mean-coordinate
equations derived in Ref.\ \onlinecite{ABZ}.

Since the equations of motion for the probability distribution are
Markovian, the equations of motion for two-time correlation
functions of the charge can be obtained from the equations for the
average quantities [Eqs. (\ref{mf1}-\ref{mf3})] using the
regression theorem.\cite{Gardiner} Thus we obtain,
\begin{eqnarray}
\frac{d}{d \tau}\langle Q_D(t+\tau)Q_D(t)\rangle&=&
-e\Gamma_R\langle Q_D(t)\rangle -\frac{V_{\rm ds}}{eR}\langle
Q_D(t+\tau)Q_D(t)\rangle-\frac{m\omega_0^2x_0}{eR}\langle
x(t+\tau)Q_D(t)\rangle \label{q1}\\
\frac{d}{d \tau}\langle x(t+\tau)Q_D(t)\rangle&=&\langle u(t+\tau)Q_D(t)\rangle \label{q2}\\
\frac{d}{d \tau}\langle u(t+\tau)
Q_D(t)\rangle&=&-\omega_0^2\langle x(t+\tau) Q_D(t)\rangle
-\frac{\omega_0^2x_0}{e}\langle Q_D(t+\tau)
Q_D(t)\rangle-\gamma_e\langle u(t+\tau) Q_D(t) \rangle \label{q3}.
\end{eqnarray}
The initial conditions (at $\tau=0$) are given by the steady state
properties of the system. The charge noise spectrum is then
obtained by integrating the coupled equations to obtain $\langle
Q_D(t+\tau)Q_D(t)\rangle$, and then performing the necessary
Fourier integral.

\begin{figure}[t]
\center{\epsfig{file=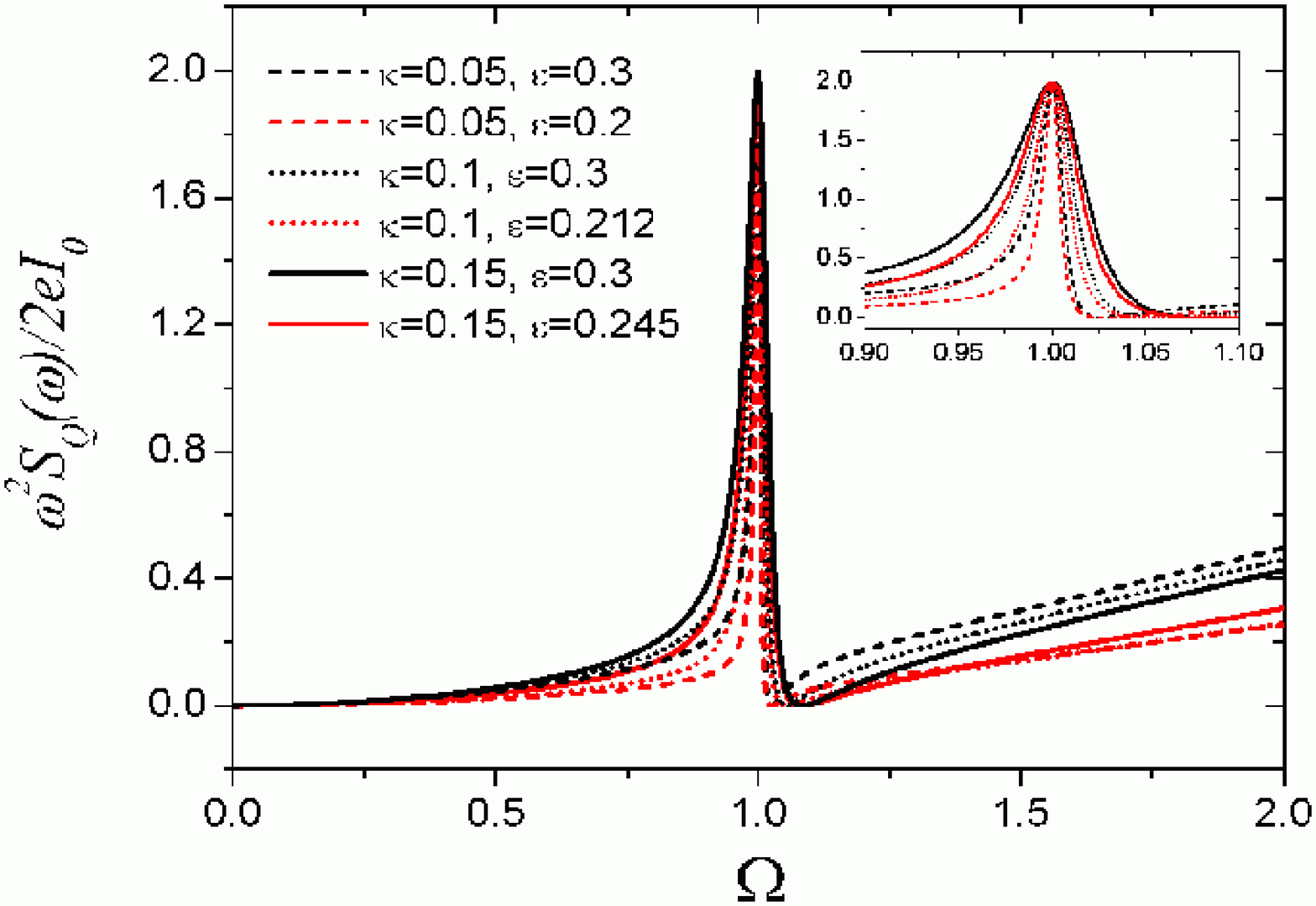,width=8.0cm} }
  \caption{Frequency scaled charge noise spectrum, $\omega^2S_Q(\omega)$, for
  a range of resonator parameters, with $\gamma_e\tau_t=\Theta=0$ and $\Gamma_L=\Gamma_R$.} \label{fig:qnoise}
\end{figure}

The frequency scaled spectrum, $\omega^2 S_Q(\omega)$, is shown in
Fig.\ \ref{fig:qnoise} for a range of resonator couplings and
frequencies. The charge noise spectrum of the SET island, like the
function $G(\omega)$ considered earlier, contains only a single
peak, at the renormalized resonator frequency. This is not
surprising as the equations of motion from which $S_Q(\omega)$ is
obtained [Eqs.\ (\ref{q1}-\ref{q3})] have a structure which is
very similar to those which determine $G(\omega)$ [Eqs.\
(\ref{one}-\ref{three})]. Figure \ref{fig:qnoise} shows that the
height of the $\Omega=1$ peak in $\omega^2 S_Q(\omega)$ is almost
independent of $\kappa$ and $\epsilon$ (provided $\kappa$ and
$\epsilon$ are much less than unity, and also non-zero), this is
because the peak height in the charge noise itself is
approximately determined by the ratio of the coupling to the
intrinsic damping
$\kappa/(\gamma_i\tau_t)=1/\epsilon^2$.\cite{foot}

\section{Full current noise spectrum}
\begin{figure}[t]
\center{\epsfig{file=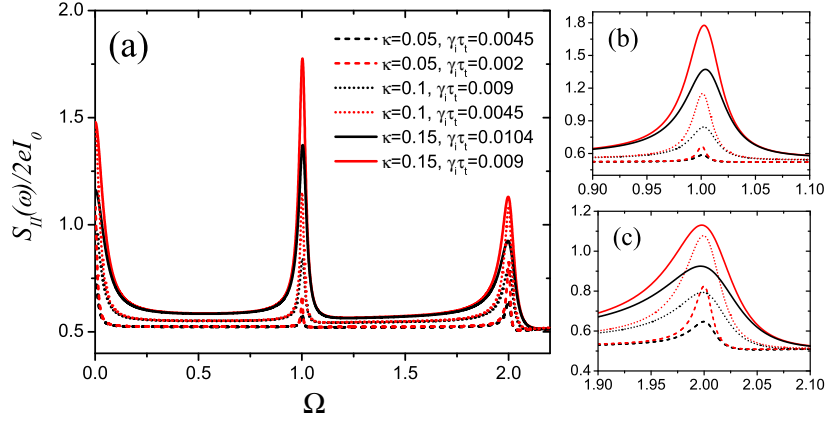,width=12.0cm} }
  \caption{Full current noise spectrum, $S_{II}(\omega)$,  for a range of resonator parameters, assuming $a=b$
  with $\Theta=0,\gamma_e\tau_t=0$. The full spectrum is shown in
  (a) while the insets (b) and (c) show the details of the peaks at $\Omega=1$
  and 2, respectively. All calculations are performed for
$\Gamma_L=\Gamma_R$.} \label{fig:full}
\end{figure}

Once the noise spectra of the tunnel currents and the charge on
the island have been calculated, the full current-noise spectrum
is readily obtained by combining its components in the appropriate
way [as given by Eq.\ (\ref{abeq})]. Figure \ref{fig:full} shows
the full current-noise spectrum for the particular choice of the
parameters $a=b$. The overall spectrum is dominated by peaks at
$\Omega=0,1$ and $2$, and largely flat elsewhere. The peak in the
charge noise spectrum suppresses the peak in the tunnel noise
spectra at $\Omega=1$, so that the peak at $\Omega=2$ can become
larger in the overall spectrum. For a combination of sufficiently
strong coupling and weak damping, the zero-frequency noise exceeds
the Poissonian value $2eI_0$.

In the limit where $\kappa$ goes to zero, the full current noise
spectrum is readily obtained analytically. In this limit the left
and right-hand junction current noise spectra are the same, and
the full current noise (spectrum assuming $a=b$) is
\begin{equation}
S_{II}^0(\omega)=eI_0\left[
1+\left(1-\frac{4\Gamma_L\Gamma_R}{\Gamma^2}
\right)\frac{\Gamma^2}{\Gamma^2+\omega^2}\right],
\label{comparison}
\end{equation}
where $\Gamma=\Gamma_L+\Gamma_R$. This is of course the usual
result for sequential tunnelling through a symmetric
double-barrier structure.\cite{korot,sun} Thus in strong contrast
to the coupled system, the current noise of the uncoupled,
symmetric ($\Gamma_R=\Gamma_L$), SET would be flat:
$S^0_{II}(\omega)/(2eI_0)=0.5$.

Figure \ref{fig:full} shows in detail the dependence of the
spectrum on the resonator coupling, $\kappa$ and intrinsic damping
$\gamma_i\tau_t=\kappa\epsilon^2$. Once the frequency of the
spectra is scaled by the renormalized frequency, a relatively
simple behavior emerges. Increasing the coupling strength
increases the height and widths of the peaks, whilst increasing
$\gamma_i$ (for fixed coupling) simply rounds the peaks down
without affecting their widths. Although the heights of the peaks
at $\Omega=1$ and 2 scale with coupling in a very similar way in
the individual tunnel current spectra (see Fig.\
\ref{fig:cnoise1}), the relative magnitudes of the peaks at
$\Omega=1$ and $\Omega=2$ in the full spectrum are very sensitive
to the magnitude of $\kappa$. This is because the peak in the
frequency-scaled charge noise $\omega^2 S_Q(\omega)$ is almost
completely insensitive to changes in $\kappa$ and hence gives rise
to a proportionately stronger suppression of the $\Omega=1$ peak
for smaller $\kappa$ values.

\begin{figure}[t]
\center{\epsfig{file=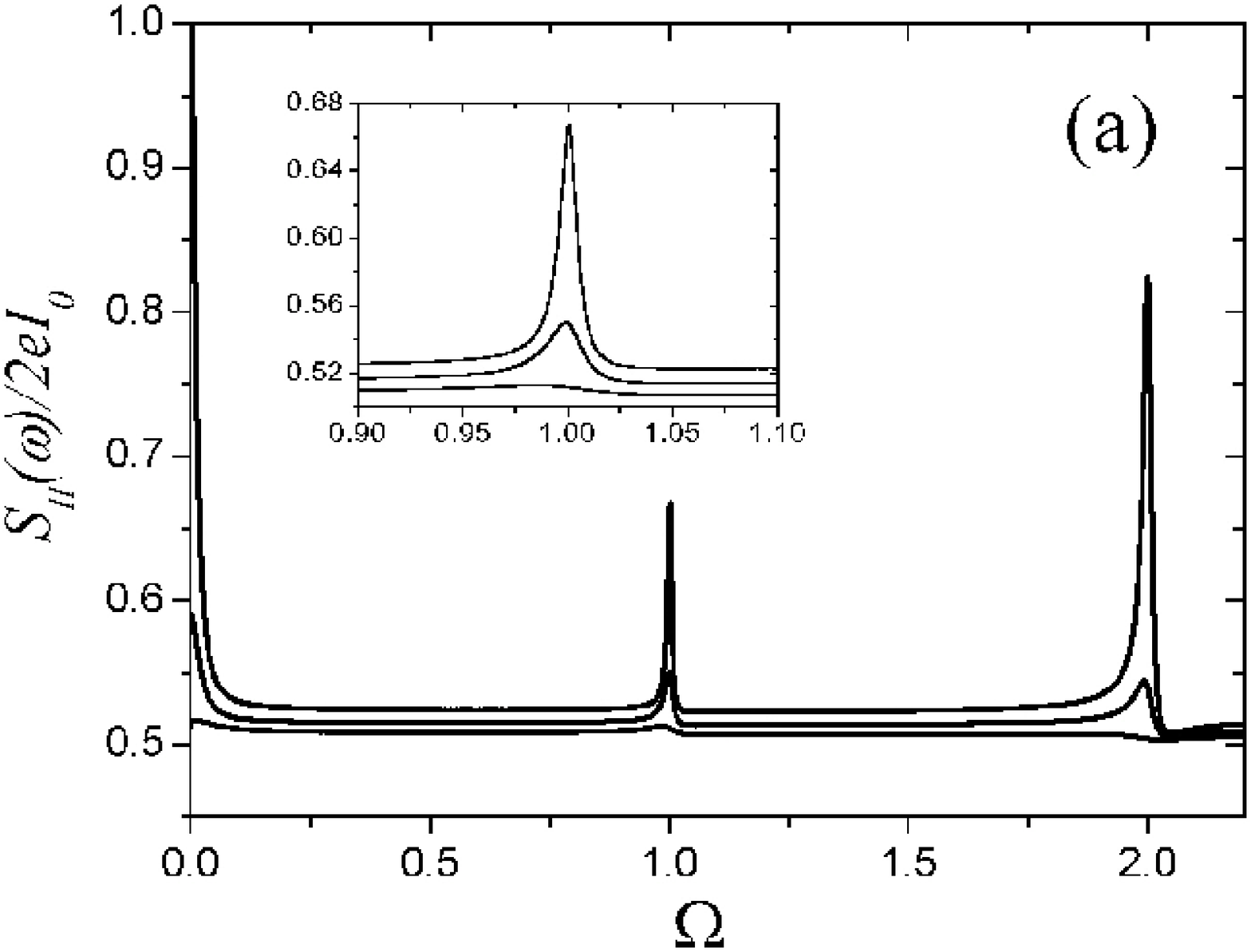,width=7.0cm}
\epsfig{file=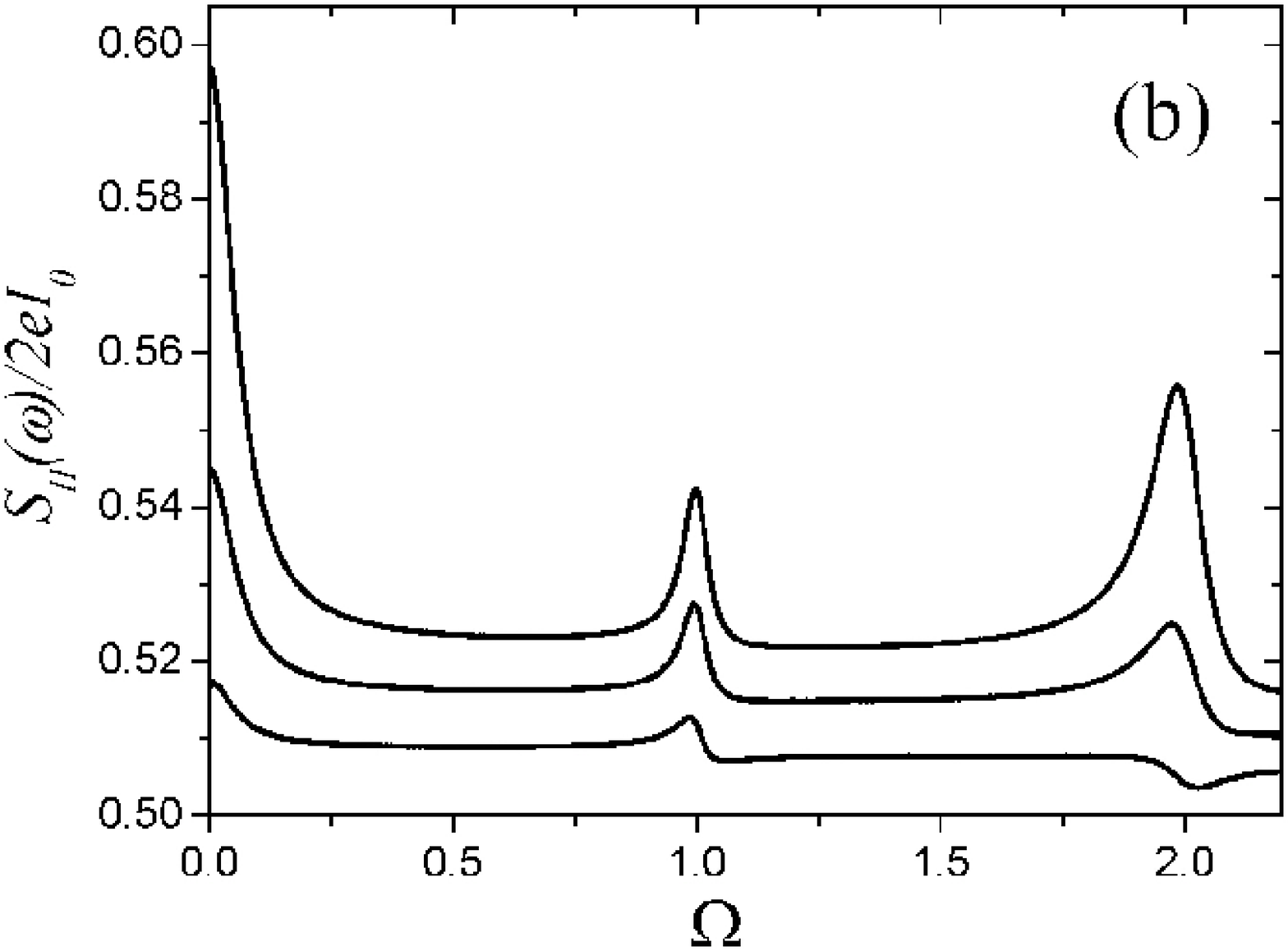,width=7.0cm} }
  \caption{Full current noise spectrum including extrinsic effects.
  The influence of extrinsic damping at $\Theta=0$ are shown in (a) where the
  curves (from top to bottom) are for  $\gamma_e=0,\gamma_i$ and
  $5\gamma_i$. The effect of a finite background temperature is shown in (b) where the
  curves (from top to bottom) are for  $\Theta=0.2,0.1$ and
  $0$, and $\gamma_e=5\gamma_i$ in each case.
   The calculations are performed for  $\kappa=0.05,\epsilon=0.2$,
 $a=b$ and $\Gamma_L=\Gamma_R$.} \label{fig:full2}
\end{figure}

For the relatively low frequencies shown in Fig.\ \ref{fig:full},
the noise is higher than the uncoupled value, even away from the
immediate vicinity of the peaks. This increase in the background
noise level depends on $\kappa$, but not $\epsilon$, and occurs
because the resonator breaks the symmetry of the tunnel rates
through the two junctions. Only for the uncoupled system do the
physical tunnel rates correspond to the parameters $\Gamma_L$ and
$\Gamma_R$. When the coupling to the resonator is switched on, the
tunnel rates for both junctions depend linearly on the position of
the resonator, but with opposite sign, and the strength of this
coupling is given by $\kappa$. The coupling to the resonator can
thus be thought of as giving rise to an effective renormalization
of the tunnel rates and since the noise is a minimum for the
symmetric tunnel rates of the uncoupled system,
$\Gamma_L=\Gamma_R$ (which we choose here), the coupling  always
gives rise to an increased background noise level. Although it is
not shown in the figures, the current noise reduces to the
uncoupled value at large frequencies, $\Omega \gg 1$, since any
asymmetry in the junction tunnel rates becomes unimportant at
large frequencies.

The influence of extrinsic effects on the current noise spectrum
 are illustrated in Fig.\ \ref{fig:full2}. As with the tunnel current spectra,
  extrinsic damping reduces the peak heights and the background noise level, whilst increasing the background
temperature has the opposite effect. Furthermore, it is
interesting to note that extrinsic effects can alter the relative
heights of the peaks at $\Omega=1$ and $2$.

The full noise spectrum could equally well be calculated for a
variety of cases where $a\neq b$, reflecting the balance of
capacitances in the system more accurately, as discussed in Sec.\
II. However, now that we have obtained the individual spectra of
the noise in the tunnel currents and the charge noise, it is clear
that for small changes of the coefficients about the values
$a=b=1/2$, there would be no qualitative change in the overall
current spectrum and so we have not investigated such cases
explicitly.

\section{Conclusions}
The presence of a voltage-gate consisting of a nanomechanical
resonator has a dramatic effect on the current-noise spectrum of
the SET.  The low frequency noise of the SET can be substantially
enhanced and two additional peaks appear in the noise spectrum: at
the frequency of the resonator (renormalized by the interaction
with the SET) and at twice that frequency. Without extrinsic
damping and with the background temperature set to zero, the peaks
in the current noise are sharp, with heights controlled by the the
strength of the resonator-SET coupling and the magnitude of the
intrinsic damping constant.

 The features in the
current-noise spectrum reflect the fact that the coupled dynamics
of the SET and the resonator lead to important correlations
between charges flowing through the SET and the position of the
resonator. In particular, the interaction between the resonator
and the electrons on the SET island provide a mechanism by which
the passage of one electron through the system can affect the
motion of subsequent electrons. Hence it is not surprising that
the noise in the measured current does not conform to any simple
picture based on a combination of the usual noise spectrum of the
SET and the mechanical noise spectrum of a damped harmonic
oscillator.

The calculation presented here is for the particular case of a
classical resonator coupled to a SET as a voltage gate. Very
recently a number of other authors have considered aspects of the
current noise in a similar nanoelectromechanical system known as
the electron shuttle,\cite{shut} in both the
classical\cite{pist,isac} and quantum regimes.\cite{novotny} In
particular, the current noise spectrum of a classical shuttle
system obtained by Isacsson and Nord\cite{isac} shows peaks at the
frequency of the mechanical system and its first harmonic in close
correspondence with the spectrum of the SET-resonator system
described here.

\section*{Acknowledgements}
It is a pleasure to thank M.P. Blencowe, A.M. Martin and D.A.
Rodrigues for a series of very useful discussions. This work was
supported by the EPSRC under grant GR/S42415/01.
\appendix
\section{Calculation of Steady-State Properties}
The moments of the steady-state distribution which appear in the
set of equations of motion describing the motion of charges
through the SET [Eqs.\ (\ref{one}-\ref{final})] can be obtained
analytically from the master equations [Eqs.\ (\ref{spn}) and
(\ref{spn1})]. Adding Eqs.\ (\ref{spn}) and (\ref{spn1}) leads to
an equation of motion for $P(x,u;t)$ which in turn can be used to
obtain an equation of motion for the moment $\langle
x^nu^k\rangle$. A similar equation of motion for the moment
$\langle x^nu^k\rangle_{N+1}$ can be obtained directly from Eq.\
(\ref{spn1}).

Assuming a steady state, the equations of motion for $\langle
x^nu^k\rangle$ and $\langle x^nu^k\rangle_{N+1}$ lead to the
following linear equations,
\begin{eqnarray}
k^2\omega_0^2\left[\langle x^{n+1}u^{k-1}\rangle-x_0\langle
x^{n}u^{k-1}\rangle_{N+1}\right]&=&n\langle u^{k+1}x^{n-1}\rangle
-k\gamma_e\langle u^kx^n\rangle+k(k-1)\frac{G}{2m^2}\langle
u^{k-2}x^n\rangle\\
k^2\omega_0^2\left[\langle x^{n+1}u^{k-1}\rangle_{N+1}-x_0\langle
x^{n}u^{k-1}\rangle_{N+1}\right]&=&n\langle
u^{k+1}x^{n-1}\rangle_{N+1} -(k\gamma_e+\Gamma_L+\Gamma_R)\langle
u^kx^n\rangle_{N+1}\\&&+k(k-1)\frac{G}{2m^2}\langle
u^{k-2}x^n\rangle_{N+1}+\Gamma_R\langle u^k
x^n\rangle+\frac{m\omega_0^2x_0}{Re^2}\langle x^{n+1}u^k\rangle,
\nonumber
\end{eqnarray}
respectively. By considering these equations for appropriate
values of $k$ and $n$, a set of linear equations is generated
which can be solved to obtain the necessary steady-state moments,
assuming only that the steady-state probability distribution is an
even function of the velocity, $u$.


\begin{thebibliography}{99}
\bibitem{neglect} This is consistent with previous treatments of
the SET which usually make the assumption that when $C_g\ll C_j$,
one can set $a=b=1/2$, as discussed in Ref.\
\onlinecite{reallyuseful}.
\bibitem{pchoice} The range of values of $\epsilon$ and $\kappa$
which might be achieved in practice is discussed in Ref.\
\onlinecite{ABZ}, but in general it is expected that SET-resonator
systems will operate in the regime where $\epsilon, \kappa\ll 1$.
\bibitem{foot} This is readily found from the approximate solution for the
charge-charge correlation function, valid
 for $\epsilon\ll 1$, that is obtained via the approach used to solve the mean coordinate
equations in Ref.\ \onlinecite{ABZ}.
\end{thebibliography}
\end{document}